\documentclass{PoS}
\usepackage{color}
\usepackage{url}

\title{Searching for Variability of the Crab Nebula Flux at TeV Energies using MAGIC Very Large Zenith Angle Observations}

\ShortTitle{Crab Nebula Variability at TeV Energies with MAGIC}

\author{\speaker{Juliane van Scherpenberg}$^a$, Razmik Mirzoyan$^a$, Ievgen Vovk $^a$, Michele Peresano $^b$, Darko Zari\'c$^c$, Petar Temnikov$^d$, Nikola Godinovi\'c$^c$, J\"urgen Besenrieder$^a$} \author{for the MAGIC Collaboration\footnote{\texttt{https://magic.mpp.mpg.de/}. For collaboration list see PoS(ICRC2019)1177} \\
\llap{$ˆa$} Max-Planck-Institute f\"ur Physik\\ D-80805 M\"unchen, Germany\\
\llap{$ˆb$} CEA-Saclay-/ Irfu-DAp / UMR AIM\\
Orme des merisiers, B\^at, 709-91191 Gif-sur-Yvette Cedex, France\\
\llap{$ˆc$} Faculty of Electrical Engineering, Mechanical Engineering and Naval Architecture\\ University of Split - FESB, 21000 Split\\
\llap{$ˆd$}  Inst. for Nucl. Research and Nucl. Energy, Bulgarian Academy of Sciences\\ BG-1784 Sofia, Bulgaria\\

E-mail: \email{jvsch@mpp.mpg.de}}

\abstract{The Crab nebula was once considered to be a stable source until strong flares, up to 30 times increase in flux, were observed in the MeV and GeV energy range by the AGILE and Fermi Gamma-ray Observatories. Existing nebula models often assume that the electron population emitting synchrotron radiation at lower energies is responsible for the VHE emission via Inverse Compton (IC) scattering. This suggests that the variability of the synchrotron $\gamma$-ray emission may also become observable in the multi-TeV energy range. Until now, no variability in the Crab Nebula flux has been found in the VHE regime.
In 2015, MAGIC started an observational campaign which improved the collection efficiency of $\gamma$-rays above several tens of TeV. These observations are performed under Very Large Zenith Angles (VLZA) and lead to a large increase in the collection area. This allows us to observe the low fluxes at TeV energies in a shorter time compared to standard observations, and to significantly increase the observable energy towards higher energies. We have studied the Crab Nebula light curve obtained from the VLZA data since 2015 in search for the flux variability at the highest TeV energies. The results of this study will be presented.
}

\FullConference{36th International Cosmic Ray Conference -ICRC2019-\\
		July 24th - August 1st, 2019\\
		Madison, WI, U.S.A.}

\begin{document}
 
\section{Introduction}
The Crab Nebula is the remnant of a supernova explosion that was observed in 1054 AD. A pulsar at its center powers a pulsar wind nebula (PWN) by constantly injecting ultra-relativistic electron and positron pairs into the surrounding environment. The electrons are randomized in the magnetic field of the nebula and emit synchrotron radiation from radio to $\gamma$-rays of $\sim100$ MeV. A second component in the photon emission spectrum ranging from few GeV to several tens of TeV $\gamma$-rays is commonly attributed to inverse Compton up-scattering of the various low energy photon fields by the electrons. For many years, the Crab Nebula has been considered to have a constant $\gamma$-ray flux and has been established as a `standard candle' for $\gamma$-ray astronomy. The Large Area Telescope (LAT) onboard the Fermi Gamma-ray Space Telescope (\textit{Fermi}) and the AGILE satellite reported on bright flares in $\gamma$-rays from the Crab Nebula for the first time in 2011 (\cite{Fermi_flares}, \cite{AGILE_flares}). These original flares were observed between 2007 and 2010 with 3-6 times higher fluxes with respect to the average at energies above 100 MeV. Since then, roughly one flare per year has been observed, the brightest ones appearing in April 2011 and March 2013 with a flux increase up to a factor 30 and 20, respectively (\cite{Fermi_flare_April2011}, \cite{Fermi_flare_March2013}).
The origin of these flares is still not completely understood. Various models follow different approaches to explain the sudden outbursts on short timescales some of which predict that the flux increase could be visible as well in the IC component of the SED at high TeV energies. The new very large zenith angle (VLZA) observational technique introduced by MAGIC offers a more effective way of detecting $\gamma$-rays above several tens of TeV. Observations of the Crab Nebula under VLZA offer the possibility to study its flux variability at the highest accessible $\gamma$-ray energies and provide more insight into the origin of the Crab Nebula flares. 
 
\section{Observations}
\subsection{The MAGIC Telescopes}
The MAGIC (Major Atmospheric Gamma Imaging Cherenkov) Telescopes are two 17m diameter Imaging Atmospheric Cherenkov Telescopes (IACTs) located at the Roque de los Muchachos Observatory on La Palma, Spain ($28^\circ$N, $18^\circ$W) at an altitude of 2200m a.s.l. The two telescopes observe $\gamma$-rays above $\sim 30$ GeV by imaging the short Cherenkov light flashes produced by the extensive air showers (EAS) that they initiate in the upper atmosphere. The EAS are observed by both telescopes in stereoscopic mode. The integral sensitivity of the MAGIC telescopes is $(0.66 \pm 0.03)\%$ of the Crab Nebula flux for E > 220GeV and 50hrs of observations at zenith angles $<30^\circ$. 
\subsection{Data Sets}
The data presented here was taken between December 2014 and April 2018 at very large zenith angles (VLZA) between $70^\circ$ and $80^\circ$. This campaign specifically aimed to increase the sensitivity of MAGIC at the highest energies. This was possible because of the large increase of the effective area that is achieved at VLZA leading to a more efficient detection of the highest energy $\gamma$-rays (E > 10TeV) which is typically limited by the low fluxes of astrophysical sources at these energies. For more details about the VLZA observation technique and its achievements see \cite{ICRC_Michele} and \cite{ICRC_Poster}.

\subsection{Data Analysis}
The obtained data was analysed with the standard MAGIC Analysis and Reconstruction Software (MARS, \cite{mars}). 
The data set was divided into a 'Rise' and a 'Set' sample depending on the azimuth angle under which the observations were performed, corresponding to the source rising/setting over the horizon. This procedure ensures that this analysis accounts for the potentially non-negligible azimuthal dependence of the stereo reconstruction quality.

\subsection{Light Curves}
We produce a light curve with a minimum energy of 10 TeV. The data is binned on a monthly time scale to ensure enough statistics in each bin. The amount of effective observation time in each of these monthly bins varies between $\sim25$min and $\sim300$min depending on the amount of scheduled observations in the repsective month and the quality of the data.

\section{Study of variability}

\subsection{Constant Fit}

To evaluate whether the light curve is variable, we fitted a constant flux to the data and evaluated the fit probability.
The Poissonian likelihood function 
\begin{equation}
 \mathcal{L}(f, N_{off}) = \mathcal{L}_{source}(N_{on} | f,N_{bkg}) + \mathcal{L}_{bkg}(N_{off} | N_{bkg})
\end{equation}
was maximized, where $\mathcal{L}_{source}$ gives the probability for the number of events $N_{on}$ that were observed in the direction of the source, given a constant source flux $f$ and the number of background events $N_{bkg}$ in the respective time bin. $\mathcal{L}_{bkg}$ is the probability for the observed number of events coming from the off-source region $N_{off}$ given $N_{bkg}$.
The value $F_{fit}$ that maximizes the likelihood is $F_{fit} = (0.25 \pm 0.03) 10^{-12} cm^{-2} s^{-1}  $. Uncertainties are statistical only.

The fit probability was determined by simulating data sets with a number of bins equal to the light curves. The number of off-source events in each bin were drawn from a Poisson distribution around the best fit value for the background in the respective bin. The Poisson parameter for the simulation of on-source events was calculated as the sum of the fitted background events and the fitted excess events, resulting from the multiplication of $F_{fit}$ with the effective area and observation time in each bin. For each simulated light curve the likelihood $\mathcal{L}_{sim}$ with respect to $F_{fit}$ was calculated. The p-value of the fit to the original light curve was obtained from the cumulative distribution $\mathcal{L}_{sim}$. The resulting fit probability is p=0.99, showing a high probability for a stable flux across the overall lightcurve.

\subsection{Upper Limits on Variability}
We calculated the fractional variation of the light curve obtained from data as well as from simulations. 

The fractional variation is defined as
\begin{equation}
 F_{var} = \frac{\sqrt{\sigma^2 - \delta^2}}{f_{fit}}
\end{equation}
where $\sigma^2$ is the variance of the data and $\delta^2$ their mean squared uncertainties.

From this, the scale of the variability that is present in the light curve is evaluated. From the distribution of $F_{var}$ for the simulated light curves we obtain an 99.7\% upper limit on the variability at 1.86.

\subsection{Sensitivity to general fluctuations}
\label{sec:sens_general}
Next, we evaluate at which scale of variability of the Crab Nebula flux MAGIC VLZA observations would be sensitive enough for a significant detection.
For this, light curves where simulated similar as described above. However, the flux parameter was not assumed to be constant. Instead in each bin the flux was multiplied with an additional random fluctuation drawn from a uniform distribution within a certain range. We evaluated at which scale of allowed fluctuations the distribution of $F_{var}$ can be significantly distinguished from the distribution of $F_{var}$ for a constant flux assumption, which was obtained in the previous step. For each pair of simulated light curves - one with a constant flux assumption, one with a variable flux - we calculated the difference between the two values of $F_{var}$ $$\Delta F_{var} = F_{var}^{variable} - F_{var}^{constant}$$ where we obtain the number of occurences with $\Delta F_{var} \leq 0$. 

The two distributions can be distinguished on a $3\sigma$ level if fluctuations up to $0.56 \cdot 10^{-12} cm^{-2} \cdot s^{-1}$ on top of the constant flux are allowed.

\subsection{Sensitivity to short term fluctuations}

Finally,we evaluated how the ability to significantly detect an increase in the flux depends on the scale of this increase and its duration. For a range of exposures between $3.6 \cdot 10^{13} cm^2 \cdot s$ and $5.4 \cdot 10^{14} cm^2 \cdot s$ (this corresponds to an observation time between 1 and 15 hours for a constant effective area of $1km^2$) a constant flux of $\Delta f\cdot f_{fit}$ was assumed for the simulation of light curves. In each two-dimensional bin in exposure-$\Delta$f space, 100000 samples were simulated drawn from a Poisson distribution around the expected number of On source events. The latter was obtained by adding the respective source fluxes $\Delta f\cdot f_{fit}$ and the average background flux obtained from the data and multiplying them with the respective exposures. 
Each distribution of simulated On counts in an exposure bin was compared with the distribution expected from $f_{fit}$ and the respective exposure. The probability for the two distributions to overlap was computed as in section \ref{sec:sens_general}. From this, the minimum observation time, that is needed to detect a certain increase in flux at a given significance, can be derived. Table \ref{tab:Sensitivity_flare} lists an overview of the obtained values. Note that MAGIC can observe the Crab at zenith angles $\leq 70^\circ$ roughly 1 hour per night, meaning that an exposure of e.g. 5 hours would translate to a total duration of the flare of $\sim 5$ days 

\begin{table*}
\label{tab:Sensitivity_flare}
\centering                       
  \begin{tabular}{ c c c }       
\hline\hline                 
$\mathrm{t_{obs}}$ [h] & Flux [$10^{-12} cm^{-2} s^{-1}$] & Flux/$\mathrm{Flux_{fit}}$   \\    
\hline

 \hline
   2.34 &  0.90 & 3.6 \\
   3.06 &  0.80 & 3.2   \\
   4.19 &  0.70 & 2.8 \\
   6.45 &  0.60 & 2.4\\
   11.73 & 0.50 & 2.0 \\
\hline                                   
\end{tabular}
\caption{Observation time needed to detect flux increase above nominal Crab nebula flux on 3$\sigma$ confidence level given the average effective area, $\mathrm{A_{eff}= 1.34km^2 }$ for VLZA  observations. Flux levels include systematic uncertainties.}     
\end{table*}

\section{Conclusions/Summary}
The Crab Nebula flux has been oberseved to be variable in the high energy band of the synchrotron emission. The origin of this variability is still unclear. Constraining the variability at the highest energy $\gamma$-ray can help understand the processes responsible for the variability. MAGIC VLZA observations have increased the observable energy range of MAGIC up to 100 TeV. The lightcurves spanning several years of data have been evaluated. No significant variability has been found. We have studied the sensitivity of our observations towards flux variability. We find that we would be able to detect fluctuations over the entire lightcurve at a scale of at least 1 to 2.25 times the fitted constant flux. Flares of several days of duration could be detected if they lead to a flux increase above 10 TeV of a factor of $\gtrsim 2$.

\acknowledgments \url{https://magic.mpp.mpg.de/acknowledgments_ICRC2019}

\end{document}